\documentclass{epl}
\usepackage{graphicx}
\newcommand\Vi{\mbox{\textit V}_{\rm I}}        % Impact Velocity
\begin{document}
\title{Bouncing or sticky droplets: impalement transitions on %superhydrophobic
 micropatterned surfaces}
 \shorttitle{Bouncing or sticky droplets: impalement transitions}
%\thanks{denis.bartolo@lps.ens.fr}
\author{Denis Bartolo\inst{1}\footnote{Electronic address: denis.bartolo@lps.ens.fr (corresponding author)}, Farid Bouamrirene\inst{1}, \'Emilie Verneuil\inst{2}, Axel Buguin\inst{2}, Pascal Silberzan\inst{2} \and S\'ebastien Moulinet\inst{1}\footnote{Electronic address: sebastien.moulinet@lps.ens.fr}}
\shortauthor{D. Bartolo {\it et al}}
\institute{
  \inst{1} Laboratoire de Physique Statistique de l'ENS, CNRS UMR 8550,24, Rue Lhomond,
75231 Paris C\'edex 05, France.\\
  \inst{2}Laboratoire de Physico-Chimie Curie, Institut Curie, CNRS UMR 168
26 rue d'Ulm 75248 Paris cedex 05, France.
}
%\pacs{47.05.Dh}{Hydrodynamics, hydrolics, hydrostatics}
\maketitle
\begin{abstract}
When a liquid drops impinges a hydrophobic rough surface it can either bounce off the surface (fakir droplets) 
or be impaled and strongly stuck on it (Wenzel droplets). 
The analysis of drop impact and quasi static ''loading'' experiments on model microfabricated surfaces  allows to clearly identify the forces hindering the impalement transitions.
A simple semi-quantitative model is proposed to account for the observed relation between the surface topography and the robustness of fakir non-wetting states. 
Motivated by potential applications in microfluidics and in the fabrication of self cleaning surfaces, we finally propose some guidelines to design robust superhydrophobic surfaces.
%Abstract: to asses the length of the letter...We investigate the stability of roughness induced non-wetting states upon drop impact dynamics. Model microstructured substrates. Increasingimpact velocity bouncing to sticky transition is observed. Transition associated with liquid impalement. Quasistatic loading and a simple model that allows to define the mechanisms hindering the impalementtransition whatever the dynamics scheme. Roughness is not the relevant parameter and guide for design strategies for efficient water repellant surfaces.
\end{abstract}
Some plants leaves and insects shells exhibit extreme hydrophobicity,
making the deposition of water drops on their surface almost
impossible~\cite{botanique}.
%When deposited, for instance on a Lotus leaf, water drops remain quasi spherical and roll off the surface rather than remaining stucked on it~\cite{lotuseffect}.
All these superhydrophobic biosurfaces share two common  features: they are
made of (or covered by) hydrophobic materials, and are structured at
the micron and sub-micron scales.

During the last decade  much effort has been devoted to design
artificial solid surfaces with comparable water-repellent
properties. Their potential applications range from lab on a chip
devices to self cleaning coating for clothes, glasses,...  The actual
strategy consist in mimicking superhydrophobic biosurfaces
designing rough substrates out of hydrophobic materials. To achieve
this goal both top-down and bottom-up approaches have been
successfully developed: chemical synthesis of fractal
surfaces~\cite{japonais}, growth of carbon nanotube
forests~\cite{lau}, deep silicon dry etching~\cite{belllabs}, see
also \cite{selfcleaning-natmat} and references therein. We briefly recall
 the paradigm to account for superhydrophobicity.
Two different wetting states can be observed on microstructured
hydrophobic surfaces: (i) Wenzel state: the liquid follows the
topography of the solid surface. Defining the surface roughness
$\zeta$ as the ratio between the total surface area over the
apparent surface area, the equilibrium contact angle of a liquid
drop is given by $\cos(\theta)=\zeta\cos(\theta_{\rm flat})$, where
$\theta_{\rm flat}$ is the Young contact angle on the flat
surface~\cite{wenzel}. (ii) fakir state: The liquid only contacts
the highest parts of the rough solid, air pockets remain trapped
between the solid and the liquid surface. Only a fraction $\phi$ of
the solid surface, corresponding to the extremities of the protrusions, is wetted by the liquid. As proposed by Cassie, a
water drop adopts a contact angle given by the weighted sum $\cos
\theta_{\rm fakir}=\phi\cos\theta_{\rm flat}-(1-\phi)$,
(Cassie-Baxter relation)~\cite{cassie}. More precisely, it has been
shown that the surface energy is minimal in the fakir (resp. Wenzel)
regime if $\zeta$ is larger (resp. smaller) than $\cos \theta_{\rm
fakir}/\cos\theta_{\rm flat}$. In other words, the rougher the
substrate, the more the fakir state is energetically favored.

\noindent The characterization of superhydrophobic surfaces is
usually restricted to equilibrium contact angle measurements.
However, it has been recently reported that: (i) the value of the
measured contact angle strongly depends on the way the droplet is deposited on the surface~\cite{lafuma,He} and (ii) droplets squeezed between two moderately
rough surfaces can undergo a
sharp and irreversible transition from a fakir to a lower contact angle Wenzel~\cite{lafuma}.
%This transition unambiguously demonstrates the existence of metastable fakir states.
Moreover, the drops do not 
only reduce their contact angle, but also increases their
contact angle hysteresis, the contact line appears to be strongly
pinned on the substrate, any self-cleaning properties is thus
definitely lost.
\begin{figure}
\centerline{\includegraphics[width=12cm]{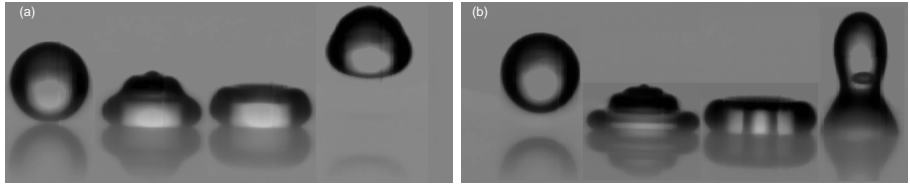}}
\caption{Rapid camera snapshots: %(a) Water drop impact on flat PDMS surface
%$\Vi=xx$,  drop radius $1 \rm mm$.
(a) Impact of a water droplet on a micropatterned surface: Pattern geometry $S_1$($h=8\,
\mu \rm m$), impact speed $\Vi=0.27\,{\rm m.s}^{-1}$, droplet radius $1\, \rm mm$. The droplet bounces off the surface. (b) Impact on the
same substrate for $\Vi=0.6\,{\rm m.s}^{-1}$, droplet radius $1\, \rm mm$. The water
droplet remains stuck on the surface. Time intervals $\approx15$ ms for the two sequences.}\label{fig:impactex}
\end{figure}
Today, a clear picture allowing to identify the mechanisms
responsible for the impalement transition from a fakir to a Wenzel
wetting state is missing. Only few theoretical explanations have
been attempted~\cite{patankar,epltho,carbone}, see also~\cite{cottin} in a different context. Beyond
the usual thermodynamic approach, it appears crucial to extend the
understanding of such impalement transition to drop impact dynamics.
Indeed almost all practical applications of superhydrophobic
surfaces rely ultimately on their ability to repel impinging drops
(rain drops, sprays,...)

In this Letter,  we present the characterization of the drop
impalement transition on microfabricated surfaces under quasi-static
and impact dynamics as well. The dual analysis of the two series of
experiments is completed by a simple model that allows to
unambiguously identify the forces hindering drop impalement and to
propose a unified criterion for the robustness of fakir non-wetting
states.

All the presented experiments have been performed on PDMS silicon
elastomer (polydimethylsiloxane, Sylgard 184 Dow-Corning) surfaces, micropatterned using
classical soft-lithography molding methods~\cite{withesides}. The
microfabricated surfaces are triangular arrays (pitch $p$) of cylindrical
pillars (radius $r$). Varying the thickness of the primary mold made of
photoresist resin (SU08, Michrochem), we have varied the pillar height from $2.7~\mu
\rm m$ up to $75~\mu \rm m$. Two different patterns have been 
used: $S_1$ ($r=11~\mu \rm m$, $p=50~\mu \rm m$) and $S_2$ ($r=9~\mu
\rm m$, $p=42~\mu \rm m$). Both patterns have the same pillars
density: $\phi\sim0.15$. Water drops lying on flat PDMS surfaces
have an advancing (resp. receding) contact angle $\theta_{\rm
a}=110^\circ$ (resp. $\theta_{\rm r}=80^\circ$). When gently deposit
on the patterned surfaces, the measured contact angles of the water drops agree with the
Cassie-Baxter relation (advancing angle $\theta_{a}\sim155^\circ$).
Fakir states are hence observed despite their surface energy would
be minimized in the impregnated Wenzel state (except for $h=75~\mu
\rm m$ pillars that should insure equilibrium in the fakir state).

A first experiment consist in studying the impact of water droplets (radius
1 mm) delivered by a precision needle on the micropatterned PDMS substrates.
Increasing the fall height increases the impact velocity $\Vi$. The
impact events have been observed using a high-speed video system (frame rate: $1000$~fps).

\begin{figure}
\centerline{\includegraphics[height=4cm]{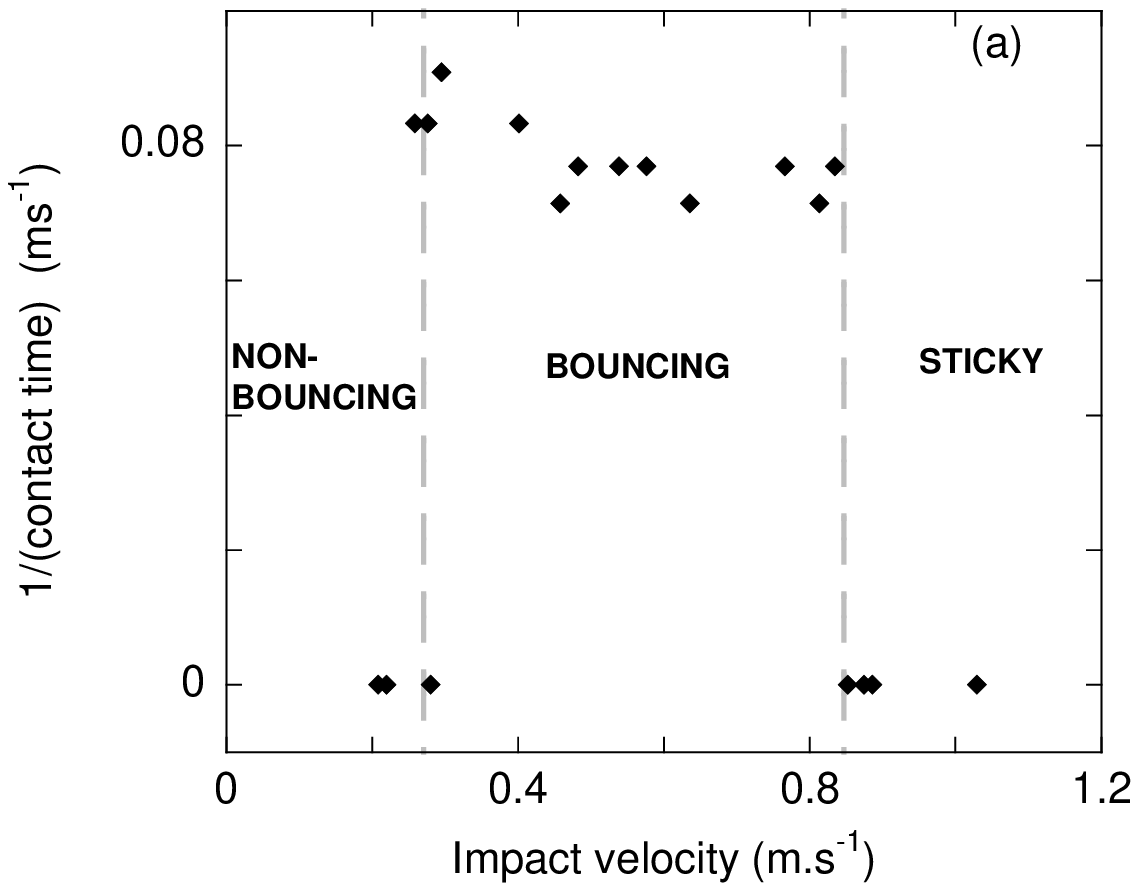}{\hspace{1cm}}\includegraphics[height=4cm]{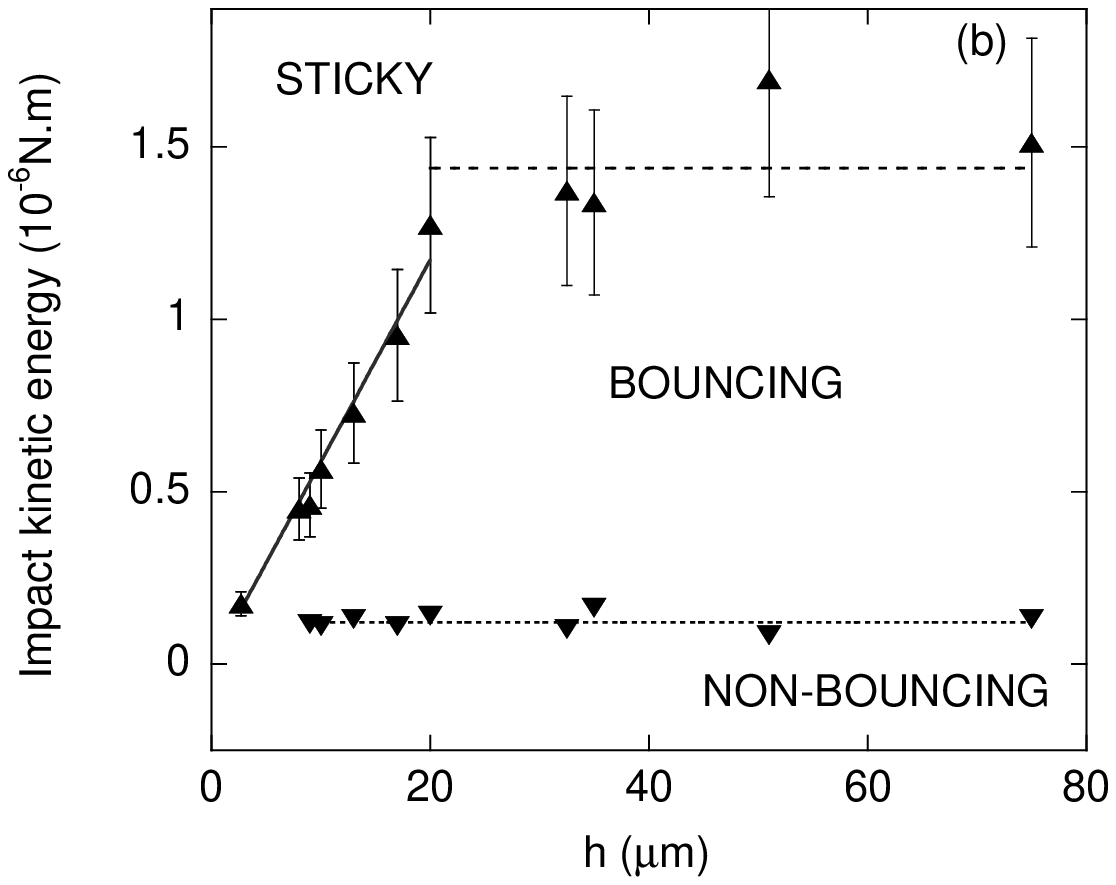}} 
\caption{(a) Inverse
of the impact contact time plotted versus the impact  velocity,
$\Vi$. Surface geometry: pattern $S_1$, pillars hight, $h=9\mu \rm
m$. (b) Critical impalement kinetic energy plotted versus pillars
height, for $S_1$ pattern geometry. $\bigtriangleup$: Critical
energies delimiting the Bouncing-Sticky transition. Plain line:
linear best fit of the data ($0<h<20\,\mu\rm m$.Dashed line: best
constant fit ($h>20\,\mu \rm m$). $\bigtriangledown$: critical
energies delimiting the bouncing-non-bouncing transition. Dotted
line: best constant fit.} \label{fig:imapactdata}
\end{figure}

The impinging drops first expand rapidly. Subsequently, due to the
hydrophobicity of the surface,  the drops retract and sometimes 
bounce off the surface as illustrated in Fig.~\ref{fig:impactex}. In
Fig.~\ref{fig:imapactdata}.a., the inverse of the contact time is
plotted versus the impact speed. We observe that bouncing occurs
only in a range of impact velocities ($V_{\rm NB}<\Vi<V_{\rm BS}$). As previously discussed by
Richard \textit{et al.} in~\cite{richardnature}, the contact time does not
depend on $V_{\rm I}$. Looking more carefully at the retraction
dynamics, three distinct regime can be identified.

-- $\Vi>V_{\rm BS}$: sticky droplets. In this regime, the contact
line hardly retracts and the intantaneous contact angle reaches
values as small as $40^\circ$, see Fig~\ref{fig:impactex}.b. This
strong pinning is a clear evidence that the microstructure has
impaled the liquid surface. Observations with a microscope have
systematically confirmed that the pillars are impregnated.

-- $V_{\rm NBB}<\Vi<V_{\rm BS}$: bouncing droplets. At intermediate
velocities, the drop bounces on the surface (this
behavior is never observed on flat PDMS), several bouncing events can
be observed. Finally, the drop remains on the surface adopting a
large contact angle consistent with the Cassie-Baxter prediction.
The initial kinetic energy of the drop is not sufficient to
overcome the energy barrier hindering the impalement
transition.

-- $\Vi<V_{\rm NBB}$:  non-bouncing droplets. As previously reported
in~\cite{richard}, we obseve a low speed threshold below which droplets do not bounce anymore. In this regime, the drop weakly expand after
impact. Then, the drop undergoes damped oscillations to reach its a quasi-spherical shape, corresponding again to a
fakir non-wetting state. Though bouncing is not observed, water does
not fill the microstructure.

Varying the height of the pillars we construct the phase diagram plotted in
Fig.~\ref{fig:imapactdata}.b. The kinetic energy thresholds
delimiting the three regimes are plotted versus $h$.

The non-bouncing to bouncing transition  can be easily understood.
During the drop retraction stage, we assume that the pinning of the
contact line is the main source of kinetic energy dissipation.
Denoting $\gamma$ the liquid-air surface tension and $\Delta\cos\theta$ the
contact angle hysteresis of the fakir drop (independent of $h$), we can assess the pinning force per unit length:
$\gamma\Delta\cos\theta$. The minimal kinetic energy needed to observe bouncing thus scales $\gamma
R^2\Delta\cos\theta$. The order of magnitude of this energy threshold is
$\sim 5\,10^{-7}{\rm N.m}$, in good agreement with what is experimentally observed.

The critical impalement threshold delimiting the bouncing and the
sticky regimes exhibits  nontrivial variations with $h$, see
Fig.~\ref{fig:imapactdata}.b. For short pillars, the critical
kinetic energy increases linearly with $h$. Above $h=20\,\mu\rm m$,
it becomes independent of the texture roughness. We insist on the irreversible nature of the impalement
transition whatever the equilibrium wetting states. This
result is one of our most important finding. 
%Thanks to the following
%set of experiments, we will positively demonstrate that the
%bouncing-sticky energy threshold give a good estimate of the energy
%barrier separating Wenzel and fakir states. Consequently, we will
%clearly show that the magnitude of this energy barrier is not a
%strictly increasing function of the surface roughness.
\begin{figure}
\centerline{\includegraphics[height=2cm]{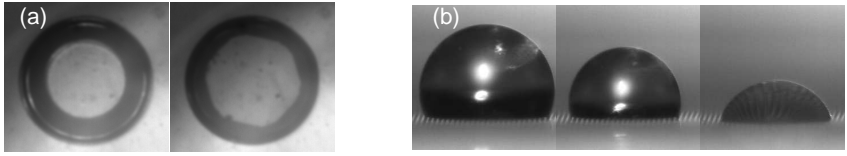}} 
\caption{(a) Two
consecutive pictures of a compressed droplet viewed from above. The
light part inside the droplet is in contact with the microstructured
substrate. Between the two snapshots the droplet has transited from
a fakir to a Wenzel state. Note that the contact line on the $S_1$
surface adopts the hexagonal symmetry of the pattern. External drop
radius(b) Three consecutive pictures taken 30 secs apart of
an evaporating droplet gently deposited on a substrate
$S_1$, ($h=20\,\mu \rm m$). Between the two last snapshots the
droplet has transited from a fakir to a Wenzel state. Contact radius of the
drop on the last picture:450\,$\mu$m}\label{fig:staticdata}
\end{figure}
We now have to identify the physical mechanism hindering the liquid
impalement upon impact of a droplet. \emph{A priori}, both capillary and
hydrodynamic forces, act on the liquid surface to impede the
impalement process. We disentangle the two effects by performing two sets of systematic quasistatic experiments. This
allows to asses to role of the sole capillary forces. 

\begin{figure}
\centerline{\includegraphics[height=6cm]{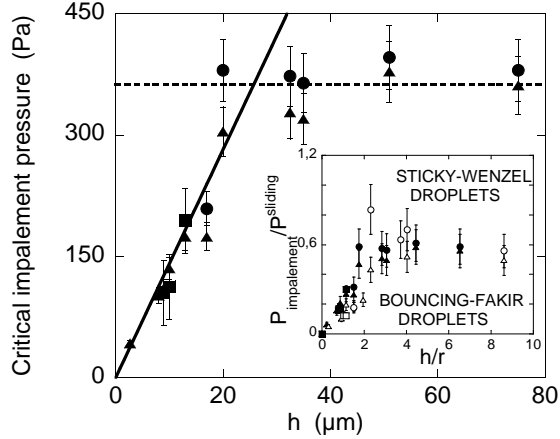}} \caption{
Impalement pressure $P_{imp}$ plotted versus the pillars'
height $h$. Circles: evaporation
experiments, squares: compression experiments, triangles: drop impact experiments. Pattern geometry: $S_1$. Plain line: best
linear fit ($h<20\,\mu\rm m$). Dashed line: best constant fit($h>20\,\mu\rm m$). Inset:
Impalement pressure normalized by $P_{\rm crit}^{\rm sliding}$, see Eq.~\ref{sliding},
plotted versus the pillars' aspect ratio $h/r$. Filled (resp. open) symbols correspond to surface $S_1$
(resp. $S_2$).}\label{fig:unified}
\end{figure}

-- Squeezing~\cite{journet}: A
droplet is squeezed between, a microstructured substrate and a
fluorinated glass slide (advancing contact angle with water
$120^\circ$). The drop is observed through the glass plate (see
Fig.~\ref{fig:staticdata}.a). The gap between the substrate and the
glass plate is slowly decreased until we observe a rapid jump
forward  of the contact line on the PDMS substrate. This jump
is the signature of the liquid impalement. The increase of the
drop Laplace pressure is the motor behind the liquid impalement.
Fitting the  droplet shape just before the transition by a surface with a constant
mean curvature ($\cal C$), we infer a critical
impalement Laplace pressure $P_{\rm imp}=\gamma{\cal C}$.
Fig.~\ref{fig:unified}. For the microstructures made of 
long pillars, our experimental setup did not allow to
reduce the gap between the two solid surfaces sufficiently to observe the impalement transition. To circumvent this technical
obstacle, we used an alternative method to increase the Laplace
pressure in the drop. 

-- Evaporation: A millimetric water droplet is gently deposited on the microstructured
substrates. Since the drop evaporates, its radius slowly
decreases with time, hence its curvature and the Laplace pressure
pushing the liquid surface on the micropillars rise continuously.
After few minutes, the drops adopts its receding contact angle
($\approx120^\circ$) and the contact line retracts. Again, above a
critical pressure $P_{\rm imp}$, one can observe a sudden variation of
the contact angle and a strong pinning of the contact line that
stops its retraction until complete evaporation, see Fig.
\ref{fig:staticdata}.b. These two observations witness the
drop impalement transition. Once a drop has reached a Wenzel state, it never relaxes toward a fakir conformation. We emphasize that arbitrarily small drops cannot be
maintained in a fakir state. This results seems to be at odd with
the the criterion: $\zeta>\cos \theta_{\rm fakir}/\cos\theta_{\rm
flat}$ which is independent of the drop size. We point out that this
criterion has been established ignoring any pressure difference
across the liquid interface, {\it i.e.} for infinitely large drops.

To quantitatively compare the outcome of our quasistatic
and impact experiments, we assess the pressure pushing the
liquid interface when a water drop hits the micropillars. Neglecting
any viscous effect, the dynamic pressure acting on the liquid
interface scales as $P\sim\frac{1}{2}\rho V_{\rm I}^2$, with $\rho$
the liquid density. Fig.~\ref{fig:unified} gathers static and impact
experimental data. The collapse
of our data on a single master curve is a strong evidence that,
hydrodynamic forces do not play any significant role in the
impalement transition on our surfaces rough at the 10~microns scale.
The wetting of the microstructure is mainly hindered by capillary
forces. This constitutes our second main results. 
%Again impalement is irreversible even on the roughest
%substrates. Whatever the impalement scheme, once a drop has reached
%a Wenzel state it never relaxes toward a fakir conformation.

\begin{figure}
\centerline{\includegraphics[height=4cm]{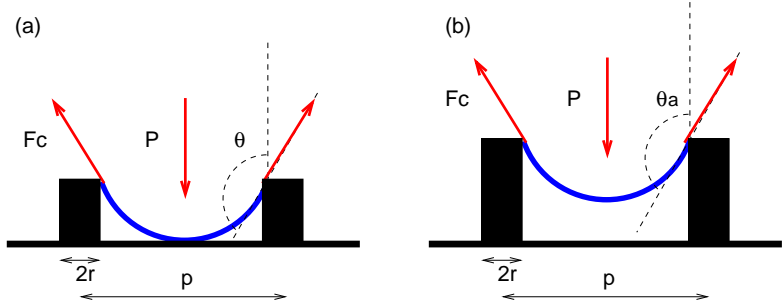}}
\caption{(a) Sketch of the ''touch down'' scenario for the fakir-sticky
droplet transition. (b) Sketch of the ''sliding'' scenario for the
fakir-sticky droplet transition }\label{fig:model}
\end{figure}

We have unambiguously identified the forces impeding the fakir to
Wenzel transition, we can now propose a simple semi-quantitative
model to account for the relation between the pattern geometry and
the critical impalement pressure. Whatever its precise origin
(curvature, liquid flow,...), the internal drop pressure
pushes the liquid interface downward. A force $F_{\rm
P}=P\left[A(1-\phi)\right]$ is experienced by the liquid free
surface enclosed by an elementary cell of the lattice, where we have
denoted $A$ the projected area of one cell. At mechanical
equilibrium, this force is balanced by the capillary force, $F_{\rm
C}=N_{\rm p} [2\pi\gamma r\cos(\theta)]$ applied at the top of
$N_{\rm p}=\phi A/{\pi r^2}$ pillars, see Fig.~\ref{fig:model}. Here
$\theta$ is the ''average'' contact angle defined on the pillars sides. Writing
explicitly the equilibrium condition $F_{\rm P}+F_{\rm C}=0$, we obtain:
\begin{equation}
P=\frac{2\phi}{1-\phi}\left |\cos(\theta)\right |\frac{\gamma}{r}.
\label{eq:equilibre}
\end{equation}
Note that in the above equation  the precise shape adopted by the
liquid interface is encoded solely in the $\cos(\theta)$ prefactor.

We now propose two impalement scenarios  characterized by two
critical values of the contact angle on the pillars.

-- ''Touch down'' scenario: increasing the drop pressure, the curvature
of the interface increases. This implies that the minimal height
separating the liquid interface and the basal surface of the
substrate diminishes. Fakir states cannot be stabilized if this
minimal height goes to zero, see Fig. 5.a. This contact condition can be expressed
in term of a critical angle value $\theta(h)$. An exact
computation of this critical angle would require the full
determination of the drop shape. To bypass this difficulty, we
can estimate this critical angle  in a much simpler manner.
Using a the small deformation approximation it is straightforward to compute profile of a fluid interface lying on top of two concentric cylinders (radius $r$ and $p$). For this simplified geometry, one can easily show that, $\cos[\theta(h)]\sim h/r$, for $r/p\ll1$ and omitting logarithmic corrections. It
then follows that the critical  ''touch down'' impalement pressure
scales at first order with respect to the solid fraction $\phi$~\cite{bicophd}:
\begin{equation}
%P_{\rm crit}^{\rm contact}\sim\frac{4\pi}{\sqrt{3}}\frac{\gamma h}{p2},
P_{\rm imp}^{\rm T}\sim\frac{\gamma h}{p^2},
\label{contact}
\end{equation}
where we have used $\phi=(2\pi/\sqrt{3})(r/p)^2$ for a triangular
lattice. Note that for denser patterns, {\it i.e.} for $2r\simeq p$, $\cos[\theta(h)]$  woulf scale as $\sim h/p$. The above equation would then be modified and $P_{\rm imp}^{\rm contact}\sim\gamma hr/p^3$. For weakly rough substrates, Eq.~\ref{contact} correctly
predicts the linear scaling of the impalement pressure with respect
to the pillars height reported in Fig.~\ref{fig:imapactdata} and
Fig.~\ref{fig:unified}. Though several simplifications were
made one can compare the estimate of the numerical
prefactor to its experimental value. For the $S_1$ surface, Eq.~\ref{contact} predicts a slope $P^{\rm contact}/h\simeq
30\,\rm Pa.\mu m^{-1}$, the slope extracted from
Fig. \ref{fig:unified} is $2$ times
smaller. We do not think that this slight difference arises solely from
the  approximations needed to estimate $\theta(h)$, a precise
numerical description of the free surface actually increases this discrepancy. 
We will show elsewhere that localized heterogeneities  
are good candidates to account for the overestimate of the $P^{\rm
contact}/h$ value~\cite{smdbnext}. Indeed, the effect of chemical or geometrical
heterogeneities cannot  be captured by our simple ''mean field'' model.

-- ''sliding'' scenario: For higher pillars a transition to a
critical impalement pressure independent of the aspect ratio $h/r$
can understood as follow. The contact angle $\theta$ has
another upper limit, namely the local advancing contact angle value
$\theta_{\rm a}$. If $\theta$ exceeds $\theta_{\rm a}$, the contact
line will spontaneously slide downward along the pillars to reach
the floor of the microstructure. Therefore, this critical ''sliding'' impalement pressure is 
obtained taking $\theta=\theta_{\rm a}$ in Eq.~\ref{eq:equilibre}:
\begin{equation}
P^{\rm S}_{\rm imp}=\frac{2\phi}{1-\phi}\left |\cos(\theta_{\rm a})\right|\frac{\gamma}{r}.
\label{sliding}
\end{equation}
Beyond the correct scaling prediction equation Eq.~\ref{sliding} provides a
rather good estimate of the experimental prefactor value for the two tested patterns, see inset in
Fig.~\ref{fig:unified}. A factor of $2$ is as good as we
could have expected given the simplicity of our model and the
precision of the microfabrication process. 

\noindent This simplified model conveys a clear picture to account
for the relation  between  the micropattern geometry and the
robustness against liquid impalement. The drop will
undergo an impalement transition if the pressure in the drop exceeds
$\min(P^{\rm T}_{\rm imp},P^{\rm S}_{\rm imp})$, the crossover between the
two impalement scenarios corresponds to an aspect ratio $h/r\sim1$ for low $\phi$ patterns.
%\sim\cos \theta_{\rm a}$.

We eventually emphasize that the above description  may be a useful
guide for the design of superhydrophobic surfaces. Efficient water repellent surfaces must obviously exhibit a high
Young contact angle, a small contact angle hysteresis and a strong
resistance against irreversible impalement. The two first
requirements can be achieved reducing the solid fraction $\phi$. The
sole comparison of the surface energies associated with Fakir and
Wenzel states would lead to the fabrication of substrates as rough
as possible. We have shown that, for a given solid fraction and
above a roughness threshold the value of the energy barrier
stabilizing the fakir states remains constant. Thus, putting efforts
to design ultra rough surface seems ineffective. Conversely,
reducing the size of the elementary pattern of the surface would
arbitrarily increases the resistance against
impalement~\cite{caillestbp}.
%Such experiments are currently under consideration.
We precise that this conclusion relies on the quantitative
agreement between our impact and quasistatic experiments.
%Howerer one could expect hydrodynamics forces to play a more important role for impacts on substrates patterns at ultra-small scales.
However, for impacts on  substrates patterned at ultra-small scales,
the confinement of the fluid flows would enhance the magnitude of
the hydrodynamic forces hindering impalement.The identification of
this characteristic scale at which they would overcome capillary
effects remains an open question.

Acknowledgments: M. Callies and D. Qu\'er\'e are gratefully acknowledged for
insightful discussions and stimulating interactions. We thank A. Ajdari, D. Bonn and J. Meunier for useful comments and suggestions. 

\end{document}